\documentclass[aps,prd,onecolumn,groupedaddress]{revtex4}
\usepackage{graphicx}
\usepackage{bm}
\usepackage{dcolumn}
\usepackage{amsmath}
\usepackage{amssymb}
\usepackage{epsfig}
\usepackage{hyperref}
\usepackage[usenames,dvipsnames]{color}
\hypersetup{
    pdfstartview={FitH},    
    colorlinks=true,       
    linkcolor=blue,          
    citecolor=blue,        
    filecolor=blue,      
    urlcolor=blue,           
    linktocpage=true
}

\def \lleq {\lower0.9ex\hbox{ $\buildrel < \over \sim$} ~}
\def \ggeq {\lower0.9ex\hbox{ $\buildrel > \over \sim$} ~}

\def \beq  {\begin{equation}}
\def \eeq  {\end{equation}}
\def \ber  {\begin{eqnarray}}
\def \eer  {\end{eqnarray}}

\newcommand{\be}{\begin{equation}}
\newcommand{\ee}{\end{equation}}
\newcommand{\ba}{\begin{eqnarray}}
\newcommand{\ea}{\end{eqnarray}}
\newcommand{\bea}{\begin{eqnarray*}}
\newcommand{\eea}{\end{eqnarray*}}

\begin{document}

\title{Why is Universe so dark ?}
\author{M. Sami}
\affiliation{Centre for theoretical physics, Jamia Millia Islamia,
New Delhi}

\begin{abstract}
In this presentation prepared for a general audience, we briefly
mention the shortcomings of standard model of universe. We then
focus on the late time inconsistency of the model dubbed age crisis
whose resolution requires the presence of a repulsive effect that
could be sourced either by dark energy or by a large scale
modification of gravity. By and large, our description is based upon
Newtonian cosmology which is simple and elegant despite of its
limitations. On heuristic grounds, we explain how a tiny mass of
graviton could account for late time cosmic acceleration. We also
include a brief discussion on the underlying physics of Type Ia
supernovae explosion and the direct confirmation of late time
acceleration of Universe by the related observations.
\end{abstract}
\pacs{}
\maketitle
\section{Introduction}
The standard model of universe dubbed {\it hot big bang} is
remarkably a successful scenario\cite{dolgov}. Its predictions about
the expansion of universe and the existence of relic radiation were
confirmed by observations several years back. Another of its success
includes the synthesis of light elements in the early universe. The
standard model relies on cosmological principle which states that
universe is homogeneous and isotropic at large scales which is
confirmed by observations. However, a homogeneous isotropic universe
evolves always again into a smooth universe. Since we have structure
in the universe, it means there should have been deviations from the
smoothness in the early universe. These tiny perturbations of
primordial nature were found by COBE in 1992. In the hot big bang
model, there is a mechanism, namely, the gravitational instability
that allows these small perturbations to grow into the structure
that we see in the universe today. However, in the standard model,
there is no way to generated these inhomogeneities. And this is
termed as one of the drawbacks of the model. In spite of the great
successes of the scenario, it seems that it has several logical in
built inconsistencies. The flatness problem: Today the observed
universe is spatially flat to a great accuracy with nearly $30\%$
dark matter content but the standard model dynamics is such that
evolution from the early times to the present universe requires ugly
fine tunings. Secondly, there are causality issues dubbed horizon
problem. The present Universe evolved from the early phase where
Universe consisted of a large number of causally disconnected
patches. Since the Cosmic Micro wave Background(CMB) is smooth to
the level of one part in $10^{-5}$, question arises, how it
happened. Since different patches in the early universe did not talk
to each other, our present universe would have shown large
patchiness in the CMB which it does not. Let us put it in the simple
language. Suppose that all of us gather here today for the first
time and we come from different Icelands which never ever interacted
by any means. It will be surprising if we happen to speak a common
language and our views on all issues turn out to be identical. There
is a remarkable  paradigm known as cosmological inflation which
beautifully addresses  the aforesaid problems. And what is most
important, it provides a quantum mechanical mechanism to generate
small perturbations that seed  structure in the universe\cite{inf}.

 The hot big bang still faces one more grave problem: The age of
universe in the model turns out to be smaller(8-10 billion years)
than the age of some well known objects(globular clusters with age in
the range, 12-15 billion years). Let us emphasize that most of the
contribution to the age comes from matter dominated universe. For
example universe was only $10^5$ years old at radiation matter
equality which is negligible compared to the age of universe.
Clearly, the age problem is related to late time expansion of
universe.
It turns out that the only way to address this problem in the
standard framework is to add some repulsive effect needed to
overcome the gravitational attraction at late times thereby giving
rise to late time cosmic acceleration. The direct confirmation of
this surprising phenomenon came from supernovae Ia observations in
1998\cite{SN} and was later supported by other indirect probes.

Late time acceleration of universe is termed as one of the most
remarkable discoveries of our times. But what causes this phenomenon
is the puzzle of modern cosmology and there is no convincing answer
to this question at present.
Normal matter (cold dark matter/radiation or baryonic matter) is
gravitationally attractive. We need an exotic matter repulsive in
character which can account for late time acceleration. The
hypothetical matter with the said unusual property is know as {\it
dark
energy}\cite{rev1,paddy,vpaddy,rev2,rev3,rev3C,rev3d,rev4,primer,Weinberg:2012es,Jain:2013wgs}.
 Cosmological constant $\Lambda$, in a sense, is the simplest! candidate for dark
 energy which is also consistent with all the observations at present. However, there
are difficult theoretical issues associated with it. Similar effect
can also be mimicked by slowly scalar rolling fields.

It is quite possible that there is no real dark energy in nature but
gravity is modified at large scales such that it reduces to Einstein
theory of general relativity in solar system where Einstein gravity
is in excellent agreement with observations and the large scale
modification gives rise to late time cosmic acceleration. Thus in
this case, the repulsive effect is provided by modification of
gravity at large scales which mimics dark energy.
It is a challenging task to build a consistent modified theory of
gravity which passes local tests and can account for late time
cosmic acceleration\cite{mirza}.

In this presentation, we give a lucid description of late time
expansion of universe using Newtonian description and try to
emphasize that the resolution of age crisis in the hot big bang
cries for late time cosmic repulsion which could be sourced by
cosmological constant/dark energy or by large scale modification of
gravity. We also describe the underlying physics of supernovae Ia
explosions.

\section{The homogeneous and isotropic universe in Newtonian description}
Cosmology is the study of structure and evolution of universe as a
whole and general theory of relativity is the valid description at
cosmological scales. In general, Einstein equations are very
complicated non linear equations but get simplified thank to the
underlying symmetries of space time. Freidmann-Robertson-Walker(FRW)
model is based upon the assumption of spatial homogeneity and
isotropy of universe known as cosmological principle. Homogeneity
means that universe looks same at any location whereas isotropy
implies indistinguishability with respect to the change of
direction. Thus homogeneity and isotropy imply that there is no
preferred location and no preferred direction in the universe.
Observations confirm the validity of cosmological principle at large
scales of the order of 100 Mpc ($1 Mpc \simeq 10^{24} cm$). In fact
this is a maximal symmetry which allows dramatic simplification in
Einstein equations. However, we would avoid here general relativity
due to conceptual and technical complications. In what follows we
would opt for Newtonian description which is simple and elegant
though has serious limitations not to be spelled out
here\cite{primer}. Using the heuristic description based upon
Newtonian approach would allow us to obtain the correct evolution
equations.
\begin{figure}[h]
\includegraphics[scale=0.5]{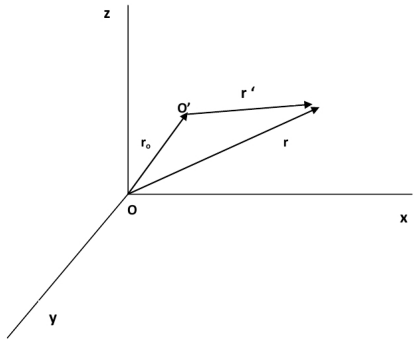}
\caption{ Figure shows a typical galaxy seen by two observers $"O"$
and $"O'"$ who try to check the validity of Hubble law in a
homogeneous and isotropic universe}
\label{hlaw}
\end{figure}
\subsection{Kinematic implications of homogeneity and isotropy}
\begin{figure}[h]
\includegraphics[scale=0.3]{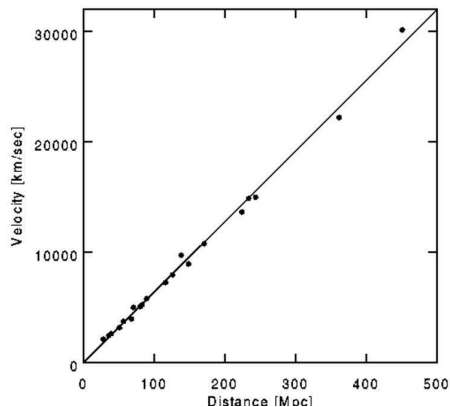}
\caption{ Hubble diagram for small  redshifts showing recession
velocities versus the redshift. Figure confirms the linear relation
between the recession velocity of a galaxies versus their distances.
}
\label{sp}
\end{figure}

Let us consider universe filled with matter in a homogeneous and
isotropic manner\cite{primer} and  imagine a coordinate system at an
arbitrary point with the origin at "O" such that the matter  is at
rest there. A priori we do not assume static universe and allow
motion of matter around the origin, the true nature of evolution
should be implied by the dynamics. A material particle with location
specified by radius vector ${{\bf r}}$(t) at time $t$ moves with
certain velocity $\bf{v}$, see Fig. \ref{hlaw}. We then ask for the
most general form of velocity distribution or velocity field
consistent with homogeneity and isotropy. It is not difficult to
guess that such a velocity distribution is given by,
\begin{equation}
\label{hubble1} {\bf v}=H(t){\bf r}(t)
\end{equation}
where $H$ is known as Hubble parameter. As for isotropy, the radius
vector transforms into a radius vector under rotations thereby an
observer looking into a different direction would see same velocity
field. Since $H$ does not depend upon $r$, homogeneity also goes
through. Indeed, let us imagine that the observer from "O" moves to
new location "O'" and observes the velocity distribution from the
new location, see Fig. \ref{hlaw}. In that case,
\begin{equation}
 {\bf r'}(t)={\bf r}(t)-{\bf r}_{O}(t)\to{\bf v'}={\bf v}-{\bf
v}_O=H({\bf r}-{\bf r}_{O})=H{\bf r'}
\end{equation}
which implies that observer located at $O'$ also sees the same
velocity distribution as the observer at "O".  Velocity distribution
given by (\ref{hubble1}) is known as Hubble law which is nothing but
the expression of homogeneity and isotropy. Hubble law is confirmed
by observations, see Fig. \ref{sp}.
 We should ,however,
remember that our description is non relativistic such that the
velocities figuring in (\ref{hubble1}) are much smaller than the
velocity of light. Hubble determined recession velocities of
galaxies by observing spectral lines of a known element emitted by
galaxies and interpreting  the result using Doppler effect. If
$\lambda_e$ and $\lambda_0$ are the emitted and observed
wavelengths, then
\begin{equation}
\frac{\lambda_0-\lambda_e}{\lambda_e}=\frac{v}{c},~(v<<c)
\end{equation}
Since all length scales in the universe scale with the scale factor,
\begin{equation}
\frac{v}{c}=\frac{\lambda_0}{\lambda_e}-1=\frac{a_0}{a(t_e)}-1\to1+z\equiv\frac{a_0}{a(t_e)}
\end{equation}
where $z$ is called redshift
 and $a_0$ designate the numerical value of the scale factor
at the present epoch when $\lambda_0$ is observed.

 Let us further explore the implications of the symmetry
expressed by (\ref{hubble1}). Formally integrating (\ref{hubble1}),
we have,
\begin{equation}
\label{com}
 r(t)={ r}_{in} e^{\int{H(t)dt}}\equiv a(t)r_{in};~~a(t)\equiv
 e^{\int{H(t)dt}}\to H=\frac{\dot{a}}{a}
\end{equation}
where $a(t)$ dubbed scale factor expresses how distances between two
points in the universe scale with time. The distance expressed by
$r(t)$ is called physical or proper distance between two points in
the expanding universe whereas $r_{in}\equiv r(t=t_{in})$ gives the
corresponding {\it comoving distance}. A comoving frame is the one
which expands with the expanding universe, matter filling the
universe appears at rest in this frame. Whole information about
dynamics of a homogeneous and isotropic universe  is imbibed in the
scale factor $a(t)$ and  we need evolution equation to determine it.

\subsection{Dynamics of Homogeneous and isotropic universe}

\begin{figure}[h]
\includegraphics[scale=0.3]{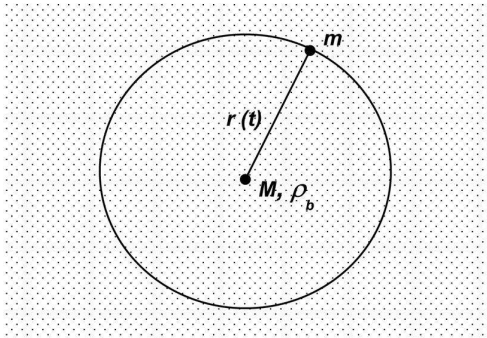}
\caption{ Figure shows a point particle with mass $m$ on the surface
of a sphere  of mass $M$ and radius $r$ in a homogeneous and
isotropic universe filled with non relativistic background matter of
density $\rho_b$. } \label{sp1}
\end{figure}

We now apply Newtonian description to FRW universe dynamics. Let us
imagine a sphere of radius $r$ and mass $M$ in a homogeneous and
isotropic universe filled with non relativistic background matter of
density $\rho_b$, see Fig. \ref{sp1}. Let us further imagine a unit
mass $m$ on the surface of the sphere and ask for the force it
experiences due to matter inside the sphere (we assume that matter
outside the sphere does not contribute to force on $m$ which is
under question in an infinite universe, see Ref.\cite{primer} for
details),
\begin{equation}
\label{aceq}
 \frac{d^2 r}{dt^2}=-\frac{GM}{r^2}\equiv -\frac{4\pi
G}{3}\rho_b(t)r(t)\to \frac{\ddot{a}}{a}=-\frac{4\pi G}{3}\rho_b(t)
\end{equation}
 In order to integrate this equation we need to know matter
density. Making use of continuity equation, we have
\begin{equation}
\frac{\partial \rho_b}{\partial t}+({\bf \nabla}.{\bf v})=0.
\end{equation}
Using then the Hubble law for velocity field, we find,
\begin{equation}
\label{rho}
 \frac{\partial \rho_b}{\partial t}+3 H\rho_b=0 \to
\rho_b=\rho_0\left(\frac{a_0}{a}\right)^3
\end{equation}
where the subscript "0" designate the corresponding quantities at
the present epoch. We can now integrate (\ref{aceq}) by substituting
$\rho_b$ from (\ref{rho}) and multiplying (\ref{aceq}) left right by
$\dot{a}$,
\begin{equation}
\label{Feq}
 H^2=\frac{8 \pi
G}{3}\rho_b-\frac{K}{a^2};~~K\equiv a_0^2\left(\frac{8\pi G
\rho_0}{3}-H_0^2\right)
\end{equation}
where $K$ is constant of integration. The redundant set of evolution
equations (\ref{aceq}),(\ref{rho}) $\&$ (\ref{Feq}) should tell us
whether universe is static($\dot{a},\ddot{a}=0) $ or evolving.
Indeed, equations (\ref{aceq}) $\&$ (\ref{Feq}) do not admit a
static solution as in this case \ref({aceq}) would imply
$\rho_b=0$(Eq.(\ref{Feq}) could be satisfied for $ K>0$). Thus
Newtonian cosmology gives rise to an evolving universe (this was
first observed in 1895-96, see Ref\cite{primer} and references
therein). This certainly went against the common perception at that
time which existed till the work of Friedmann.

 Let us note that our
description is valid for non relativistic matter at scales much
smaller than the Hubble scale. The equation of continuity
(\ref{rho}) is not applicable to relativistic matter which has non
zero pressure. For instance, in case of thermal radiation ,
$p_{rad}=\rho_{rad}/3$. The number density of photon in the
radiation redshifts as $a^{-3}$ and since any distance in FRW
universe scales proportional to the scale factor $a$, energy of each
photon ($hc/\lambda$) would scale as $a^{-1}$ which tells us that
$\rho_{rad}\sim 1/a^4$. It is therefore clear that at early epochs,
radiation dominated over the cold dark matter. Obviously, radiation
density would not satisfy the continuity equation (\ref{rho}).
Pressure is relativistic effect which corrects energy density in
general theory of relativity. The correct continuity equation reads
as
\begin{equation}
\label{rho2}
 \frac{\partial \rho}{\partial t}+3 H(\rho+p)=0 \to \rho_{rad} \sim
 \frac{1}{a^4}
\end{equation}
which gives the right behavior for the energy density of radiation.

As mentioned earlier, our framework does not apply to relativistic
fluid which has non zero pressure. Can we understand Eq.(\ref{rho2})
despite our limitations? There is a way out using thermodynamic
considerations. Thermodynamics is a great science which applies to
any system, be it relativistic or non-relativistic, classical or
quantum$-$ {\it thermodynamic description is universal}. Let us
consider universe field by matter with energy density $\rho$ and
pressure $p$ and imagine a unit comoving volume(the corresponding
physical volume or proper volume is $4\pi a^3/3$) in the expanding
universe. Assuming the expansion to be adiabatic, the first law of
thermodynamics tells that
\begin{equation}
\label{th1}
 dE + p  dV = 0
 \end{equation}
Expressing energy density of the fluid  through its mass density, we
have,
\begin{equation}
\label{E}
 E = \frac{4 \pi}{ 3} a^3\rho c^2
\end{equation}
 Substituting
(\ref{E}) into (\ref{th1}), we obtain the continuity equation in the
expanding universe,
\begin{equation}
 \frac{\partial \rho}{\partial t} + 3 H (\rho + \frac{p} { c^2}) = 0
 \end{equation}
 We thus get the pressure correction to energy density expressed by
 Eq.(\ref{rho2}) where we used the system of units with $c=1$ (we shall adhere to this system of units
 but would reinstate $c$ whenever needed for clarity). It is
 not surprising that the modified continuity equation applies to any
 fluid, be it relativistic such as radiation or non relativistic
 like cold dark matter with zero pressure.

\subsection{Age problem in hot big bang}
Since we are working in Newtonian approximation, the matter density
in (\ref{Feq}) refers to cold dark matter such that $\rho\sim
a^{-3}$. Bringing in radiation density does not change the
estimate(see discussion in the subsection E). Secondly, for
simplicity, we assume here, $K=0$(invoking cosmology with non zero
$K$ also does not change the estimate) . In that case, Freidmann
equation tells us that $\dot{a}(t)\sim a^{-1/2}$ which readily
integrates to,
\begin{equation}
 a(t) \sim t^{2/3}\to t=\frac{2}{3}\frac{1}{H}
\end{equation}
and specializing to the present epoch we get the age of universe
$t_0$,
\begin{equation}
\label{ageest}
 t_0=\frac{2}{3}\frac{1}{H_0}
\end{equation}
depending upon the observational uncertainties, $H_0^{-1}$ varies
from 12 to 15 billion years\cite{rev4} which is a very comfortable
number as there are objects in the universe which have age in this
range. However, the factor $2/3$ in (\ref{ageest}) spoils the
estimate. Let us note that we have matter density in our
description. However, bringing in the radiation density does not
change the estimate; as mentioned in the aforesaid, universe was
only $10^5$ years old at radiation matter equality. It will become
clear in the discussion to follow that there is negligible
contribution to age from early times when radiation dominated such
that early universe modifications do not affect the age of universe.
The age crisis is generically related to late time evolution and
therefore one needs to correct the late time dynamics to address
this problem. Before we address the age issue in the standard model,
we would like to mention the broad features of steady state
cosmology which is free from the said problem.
\subsection{Steady state theory}
 The standard model of universe is based upon the {\it
imperfect} cosmological principle. One should wonder why universe
should look same at all locations and in any directions only and why
should'nt it look same at all the times also? {\it a la perfect
cosmological principle}. Why space and time should not be treated at
the same footing at cosmological scales in adherence to the basic
principles of relativity? The theory based upon the perfect
cosmological principle dubbed steady state theory would have been
aesthetically far superior than the standard model of universe.
It is interesting to note that the nineteenth century  materialist
philosophy$-$  view on the genesis of universe was based upon a
similar ideology which can be found in the classic work  by
Frederick Engels, "Dialectics of nature". According to dialectical
materialism, Universe is infinite, had no beginning, no end and
always appears same thereby leaving no place for God in such a
universe.

Let us examine the broad features of the steady state cosmology. In
this theory, the global properties of universe such as Hubble
parameter and average density of Universe should remain constant. In
that case the Hubble law readily integrates to
\begin{equation}
\label{nar}
 r(t)\sim e^{H_0 t}
\end{equation}
If we now imagine a spherical region in such a Universe, its volume
would increase exponentially ($V\sim e^{3H_0t}$) which would change
the matter density in the universe. In order to keep it constant in
adherence to the perfect cosmological principle, we would then need
to continuously create matter. Let $m(t)$ be the mass created at
time $t$. Then the mass of matter needed to be created per second
per unit volume is given by,
\begin{equation}
\frac{\dot{m}(t)}{V}=3H_0\rho_0\simeq 10^{-27} kg~ km^{-3}~year
^{-1}
\end{equation}
which implies creation of one hydrogen atom per year per unit
volume. Let us note that a steady state universe in infinitely old
($r(t)\to 0$ for $t\to -\infty$) thereby it has no age problem.
Secondly Universe is naturally accelerating in the steady state
theory by virtue of perfect cosmological principle. Unfortunately,
the theory is plagued with a grave problems related to
thermalization of the microwave background radiation. However, the
generalized steady state theory known as "Quasi Steady State
Cosmology" (QSSC) formulated by Hoyle, Burbidge and Narlikar claims
to explain the microwave background radiation as well as derive its
present temperature which the big bang cannot do\cite{jvnb}.
\subsection{Cosmological constant: Newton's law of gravitation should be complemented by Hook's law}
Since Newtonian cosmology gives rise to an evolving universe, there
was an effort to modify the Newton's law of gravitation such that
local physics is left intact and static universe is realized. Let us
write down the force on a unit mass on the surface of the
sphere\cite{primer},
\begin{equation}
\label{force}
 {\bf F}=-\frac{4\pi G}{3}\rho_b {\bf r}
\end{equation}
which implies that acceleration can be made zero provided we add to
(\ref{force}) a term proportional to ${\bf r}$ vector with positive
constant of proportionality(which we denote by $\Lambda$) thereby
complementing Newton's law of gravitation by Hook's law,
\begin{equation}
{\bf F}=-\frac{4\pi G}{3}\rho_b {\bf r}+\frac{\Lambda}{3}{\bf r}
\end{equation}
which in terms of the scale factor can be cast as,

\begin{equation}
\label{aceq2}
 \frac{\ddot{a}}{a}=-\frac{4\pi
G}{3}\rho_b+\frac{\Lambda}{3}
\end{equation}
Eq.(\ref{aceq2}) readily integrates to yield the Friedmann equation,
\begin{equation}
\label{Fl}
 H^2=\frac{8 \pi G}{3}\rho_b-\frac{K}{a^2}+\frac{\Lambda}{3}
\end{equation}
It should be noted that unlike the normal matter, the positive
cosmological constant gives rise to a repulsive effect. Similar
effect can also be mimicked by scalar fields. These systems can be
thought as an ideal fluid, for instance, cosmological constant
represents a fluid with energy density,
$\rho_\Lambda=\frac{\Lambda}{8\pi G}$. By definition, a fluid which
gives rise to a repulsive effect is known as {\it dark energy}. Let
us point out that $\rho_{\Lambda}$ does not satisfy (\ref{rho}), it
rather satisfy the modified continuity equation (\ref{rho2}) which
tells us that $p_{\Lambda}=-\rho_{\Lambda}$. Cosmological constant
represents an exotic fluid with large negative pressure which
characterizes dark energy. As mentioned before slowly rolling scalar
fields also exhibit  same behavior. Such a fluid turn gravity into a
repulsive force. Cosmological constant is a relativistic object which
we some how captured in the Newtonian framework. It is interesting
to note that,
\begin{equation}
\label{p}
 \frac{\ddot{a}}{a}=-\frac{4\pi G}{3}\rho_b+\frac{\Lambda}{3}\to
\frac{\ddot{a}}{a}=-\frac{4\pi G}{3}\rho_b-\frac{4\pi
G}{3}(\rho_{\Lambda}+3p_{\Lambda})
\end{equation}
which is also true for any relativistic fluid with energy density
$\rho_r$ and $p_r$ present in the universe(in that case,
$\rho_{\Lambda}+3p_{\Lambda}$ is replaced by $\rho_r+3p_r$. The
second term in (\ref{p}) then gives rise to acceleration provided
that $p_r/\rho_r<-1/3$, this is what we mean by large negative
pressure). In general, in presence of a relativistic fluid,  it is
clear from Eq.(\ref{Fl}) that no pressure correction should occur in
the Freidmann equation. As for the acceleration equation, it can be
obtained by using the Freidmann equation and the modified continuity
equation (\ref{rho2}). Once we incorporate pressure correction, the
evolution equations become identical to those obtained in
Freidmann-Robertson-Walker (FRW) cosmology in the frame work of
general theory of relativity with $K=0,\pm 1$. Let us also note that
the evolutions equations in case of $K=0$ exhibit a symmetry under
$a(t)\to \alpha a(t)$ where $\alpha$  is constant, thank to which we
can normalize the scale factor to a priori given value at a given
time, for instance, $a(t=t_0)\equiv a_0=1$ at the present epoch;
otherwise it would depend upon the matter content in the Universe.

Cosmological constant was first introduced in 1895-96 within the
framework of Newtonian description with a hope to get a static
universe\cite{primer}. Similar attempt, within the frame work of
general theory of relativity, was made by Einstein much later.
Unfortunately, the static universe turned out to be an unstable
solution. Indeed, the static solution ($\ddot{a},\dot{a}=0$) is
possible for a particular value of $\Lambda=\Lambda_c=4\pi G\rho_0$,
see Eg.(\ref{aceq}). Let us now perturb around the static solution,
$a(t)\to a(t)+\delta a\left(t\right)$. It is not difficult to
observe using Eq.(\ref{aceq2}) that,
\begin{equation}
\label{peq}
 \ddot{\delta}a\left(t\right) = C\delta
a\left(t\right)\to \delta a\sim e^{Ct}~ (\text{for large
t}),~~C=\frac{4\pi G \rho_0}{3}
\end{equation}
which means that the static solution is unstable. Einstein then
suggested to drop the cosmological constant from his equations. But
this is not easy, if we do it at classical level, it would come back
to us through quantum corrections. Let us retain it; it could
address the age problem of the hot big bang which seems to be the
only known solution within the standard framework. In 1998, it would
turn out to be a blessing for the standard model.

Let us note that the integration constant occurring in
Eq.(\ref{Feq}) can be zero, positive or negative. However, the
recent Wilkinson Microwave Anisotropy Probe  in the cosmic micro
wave background made it clear that $K=0$ to a very good accuracy
which means that we live in a critical Universe with density,
$\rho_0=\frac{3H_0^2}{8\pi G}\equiv \rho_{cr}$ and we would adhere
to same in the discussion to follow.
 The
criticality of universe fixes the total energy budget of universe.
The study of large scale structure reveals that nearly 30 percent of
the total cosmic budget is contributed by dark matter. Then there is
a deficit of almost 70 percent and the supernovae observations tell
us that the missing component is an exotic form of energy which turn
gravity repulsive. The idea that universe is in the state of
acceleration at present is now considered to be established in
modern cosmology.

\subsection{Cosmological constant can make the universe older}
 Let us go back to the age problem. We said that the factor $2/3$
in Eq.(\ref{ageest})  spoils the estimate. Let us show that problem
could be circumvented by introducing a repulsive effect. In early
times universe is hotter and its constituents run away with large
velocities, the role of gravity is such that  it decelerates this
motion. Let us suppose for a while that we can ignore gravity. In
that case, $v=const$, Hubble law then implies that,
\begin{equation}
t=\frac{1}{H}\to t_0=H_0^{-1}
\end{equation}
which is the correct estimate for the age of universe. But we have
matter in the universe which is attractive in nature and causes
deceleration of expansion. More is the matter density, less time is
required to reach a given expansion rate, in particular the present
Hubble rate thereby giving rise to smaller age of universe. It is
therefore clear that most of the contribution to age comes from late
times. In the early universe, radiation dominated, its energy
density was large thereby a negligible contribution to the age of
universe. Indeed, the age of universe at radiation matter equality,
was just $10^5$ years. Thus it is sufficient to consider the matter
dominant universe for the estimation of age. Further, in the matter
dominated regime, more and more contribution comes from later and
later stages of expansion\cite{primer,mirza}.

At present, there is around $30\%$ matter in the universe  and we
can not ignore gravity. To  put it in layman's language, let us
consider two trains which are running with a speed of 100 km/h. If
we wish them to reduce their speed to 50km/h, the train with the
superior breaks will take less time to achieve the said speed than
the one with inferior breaks. The presence of matter plays the role
of gravitational breaks with regard to the expansion of universe.
How can we  make the gravitational breaks inferior and improve the
age of universe? The only known way out in the standard model to
address this problem is provided by  a repulsive effect which could
encounter the gravitational attraction. The presence of cosmological
constant is equivalent to a repulsive effect and it should increases
the age of universe. Indeed let us rewrite the Friedmann equation in
the following convenient form,
\begin{equation}
H^2=H^2_0\left[\Omega_m(\frac{a_0}{a})^3+\Omega_\Lambda\right],~~\Omega_m=\frac{\rho_b}{\rho_{cr}};
\Omega_\Lambda=\frac{\rho_\Lambda}{\rho_{cr}}
\end{equation}
which can be easily integrated to obtain the age of universe,

\begin{equation}
t_0=\frac{1}{H_0}\int_0^\infty{\frac{dz}{(1+z)\sqrt{\Omega_m(1+z)^3+\Omega_{\Lambda}}}}
=\frac{2}{3}\frac{1}{\Omega_{\Lambda}^{1/2}}\ln\left(
\frac{1+\Omega_{\Lambda}^{1/2}}{\Omega_m^{1/2}}\right)
\end{equation}
where  $a_0/a\equiv(1+z)$. The variable $z$ known as redshift
quantifies the effect of expansion. For observed values of density
parameters ($\Omega_m\simeq 0.3$ $\&$ $\Omega_\Lambda\simeq 0.7$),
we find that $t_0H_0 \simeq 1$ which is not surprising as we have
dominant repulsive effect represented by the cosmological constant.
As mentioned in the introduction, the early time inconsistencies of
the standard model of universe can be taken care off by inflation.
This is really amazing that resolution of late time problem of the
model also requires accelerated expansion of universe. Thus, the hot
big bang model should be sandwiched between two phases of
acceleration: Inflation at early epochs and  cosmic acceleration at
late times.

\section{Repulsive effect from large scale modification of gravity}
Since we are not using Einstein equations, our discussion is of
heuristic nature. The main effort here is to convey the underlying
ideas. In the preceding description, we tried to capture the late
time cosmic acceleration by introducing a dark energy component
which was for simplicity represented by cosmological constant. There
is an alternative thinking in cosmology that gravity is modified at
large scales which causes repulsion responsible for late time cosmic
acceleration. Several schemes of large scale modification have been
investigated in the literature in the recent years. Amongst them,
massive gravity sounds more promising though it has not yet given a
satisfactory result in cosmology. Let us try to understand the
notion of mass in gravity. We better understand electromagnetic
interactions where the force between two charged particles situated
at a distance is due to the exchange of photons with zero rest mass
thank to which the force is of long range character. The Newtonian
force between two masses has similar behavior and the particles
dubbed gravitons which are exchanged in gravitational interaction
are also massless. Roughly speaking, if we assume that graviton has
a tiny non zero mass, the effect of mass would be felt at large
distances. The effect of large mass would be felt at small scales,
say in the solar system which is untenable as Newton's law provides
an accurate description of physics there. The size of the observed
universe is given $H_0^{-1}$ which corresponds to mass scale of the
order of $10^{-33}$ eV. At the onset such a mass scale should be
safe for the local physics where Newton's law is in good agreement
with observations. Let us write down the gravitational potential of
a massive body with mass $M$ at a distance $r$ from the body in case
the graviton has a tiny mass $m$,
\begin{equation}
\Phi_m=-\frac{GM}{r}e^{-mr},~m\sim H_0
\end{equation}
which clearly reduces to Newton's potential at small scales.
However, at large scales($r\sim R_H$ such that $mr\sim 1$), the
exponential factor or the Yukawa suppression becomes operational
giving rise to the weakening of gravity. In the standard picture
presented in the previous sections, the latter can happen only due
to a repulsive effect {\it a la } cosmological constant. It is
remarkable that cosmological constant gets linked to the mass of a
fundamental particle, the graviton which is a novel perspective.
This served as one of the motivations for the formulation of non
linear massive gravity\cite{DR,Hint,mirza}.
\subsection{Relevant scales and number estimates}
In order to have a feeling about the scales involved, let us
remember that the size of our solar system is about $10^{14} cm$ and
the size of our galaxy is around $10^{22} cm\equiv 10^4 pc$ (Parsec
(pc) is the convenient unit for distance used in cosmology).
Galaxies form clusters containing a large number of galaxies, for
instance, Coma cluster contains around 1000 galaxies. Clusters then
for super clusters such that clusters in them are joined by filament
like structures with voids in between them as large as 50 Mpc ($1~
Mpc=10^6 pc$). Universe appears smooth beyond 100 Mpc. Universe
really is clumsy at small scales and consists of very rich structure
of galaxies, local group of galaxies, clusters of galaxies, super
clusters and voids. These structures typically range from $1~ kpc$
to $100~Mpc$. The study of large scale structures in the universe
shows no evidence of new structures at scales larger than 100 Mpc.

The Hubble parameter sets important scale(s) in the universe. As we
know that the age of universe $t_0$ is given by, $t_0 \sim
H_0^{-1}$. One can also ask for the time photon traveled since the
big bang till the present epoch known as the visible size of the
universe designated by $R_H$. Strictly speaking, photons can only
travel to us from the so called last scattering surface. However,
the universe was very young at that time and bringing in the last
scattering surface would not change our estimate for $R_H$,
\begin{equation}
R_H=ct_0\simeq\frac{c}{H_0}\simeq 10^{28}~cm\equiv 10^4~Mpc
\end{equation}
Next let us consider the mass scale associated with the cosmological
constant. Using Friedmann equation(switching off matter density), we
have,
\begin{equation}
\rho_\Lambda\equiv\frac{\Lambda}{8\pi G}\simeq M_p^2H^2_0\to
M^2_\Lambda\equiv \rho_\Lambda^{1/2}\simeq M_pH_0\simeq
(10^{-3}eV)^2
\end{equation}
which should not be confused with the mass of graviton which is of
the order of $H_0\simeq 10^{-33} eV$ and in case dark energy is
described by a slowly rolling scalar field, its mass should also be
of this order. It is also interesting to find the corresponding
distance scale,
\begin{equation}
c\Lambda^{-1/2}\simeq \frac{c}{H_0}=R_H
\end{equation}
Thus the cosmological constant sets the largest distance scale or
the smallest energy scale in the universe. It is interesting that
the associated mass scale is of the order of the mass of neutrino.
Can we probe it in the laboratory ? Since, to the best of our
knowledge, dark energy does not directly interact with matter, it
can only be felt through its gravitational impact. In order to check
for its local influence, let us estimate the Newtonian acceleration
caused by the cosmological constant at the present epoch. Using
Eq.(\ref{aceq}), we have,

\begin{equation}
\label{Nac}
 \text{Newtonian acceleration}=\frac{4\pi
G}{3}r\left(-\rho_b^0+2\rho_\Lambda\right)
\end{equation}
Using Eq.({\ref{Nac}), we can have crude idea about the local
influence of cosmological constant(rigorously speaking, our
framework should not apply for small scales). The acceleration
caused by $\Lambda$ is negligible at small scales as compared to the
deceleration due to matter, for instance, $\rho_b^0\sim 10^{-24}
g/cc$ in solar system much smaller than $\rho_\Lambda\simeq
10^{-29}g/cc$ and the contribution of $\Lambda$ can safely be
ignored. However, situation changes crucially at large scales of the
order of $R_H$ where $\rho_m^0$ and $\rho_\Lambda$ are of the same
order of magnitude,
\begin{equation}
\label{Nac1}
 \text{Newtonian acceleration}=
\frac{H_0^2}{2}\left(-\Omega_m+2\Omega_\Lambda\right)R_H
\end{equation}
Indeed, given the present values of the dimensionless density
parameters ($\Omega_m\simeq 0.3;\Omega_\Lambda\simeq 0.7$), the
quantity in the parenthesis on the right hand side in
Eq.(\ref{Nac1}) is of the order of one thereby the Newtonian
acceleration becomes sizable($R_H\simeq 10^{28} cm, H_0\simeq
10^{-18}/s$); it is of the order of $10^{-9} m/s^2$. Hence, the
influence of cosmological constant is important at cosmological
scales which justifies the designation given to this constant.\\
Cosmological constant sets the lowest fundamental mass scale in the
universe. Is there a fundamental mass scale that corresponds to
highest energy scale ? The mass scale below which we can trust
classical description of gravity is known as Planck scale such that
the Planck mass $M_p\simeq 10^{18} GeV$.
\section{Cosmic acceleration and its observational confirmation}
In the preceding discussion, we argued that the known mechanism to
resolve the age crisis in hot big bang is provided by the
introduction of a repulsive effect which is naturally mimicked by
cosmological constant. Let us note that in the framework of Einstein
gravity, cosmological constant  does not require ad hoc assumption
for its introduction. It is always there, indeed, it is the other
way around that we need a theoretical justification if we want to
drop it out. At low energies there no known symmetry that would
allow us to do that. So, all is well, we need cosmological constant
to address a grave problem of the standard model of Universe and it
is there in Einstein theory. However, the agreement related to age of
Universe is an indirect one and we need the direct confirmation of
cosmic acceleration which was provided by the supernovae Ia
observations in  1998.
\subsection{Luminosity distance and cosmic history}
The distance to an object in the universe can be found out by
knowing the absolute luminosity $L$ and  the  apparent or relative
brightness $B$. In static universe, the amount of energy received
per unit area from an object of Luminosity $L$ at a distance $d$
from the source is given by,
\begin{equation}
\label{d}
 B=\frac{L}{4\pi d^2}
\end{equation}
Thus knowing $L$ and measuring $B$, one can find out the distance
from the source and this is how astronomers determine the distance of
luminous objects in the universe. Obviously, one should identify the
objects whose intrinsic luminosity is known, which are called {\it
standard candles}. Secondly, one should find out the generalization
of (\ref{d}) in the expanding universe. There are two effects which
gives rise to the decrease in $B$, the redshift of each photon and
the difference in observed rate as compared to the emission rate. It
is easier to compute the desired quantity in comoving frame as universe looks static
in this frame. Let us venture a bit
into relativity. In Minkowski space time, the distance between two
space time points is defined as,
\begin{equation}
ds^2=-c^2dt^2+dx^2+dy^2+dz^2
\end{equation}
In an expanding universe, its obvious analog is,
\begin{equation}
ds^2=-c^2dt^2+a^2(dx^2+dy^2+dz^2)=-c^2dt^2+a^2(r^2dr^2+r^2(d\theta^2+\sin^2\theta
d\phi^2))
\end{equation}
where (x,y,z) or $(r,\theta,\phi)$ are the comoving coordinates and
$t$ is the cosmic time. We can define, $d\eta=dt/a$ dubbed conformal
time such that,
\begin{equation}
ds^2=a^2(-c^2d\eta^2+r^2dr^2+r^2(d\theta^2+\sin^2\theta d\phi^2)
\end{equation}
which is the FRW metric for spatially flat geometry(K=0). Computing
luminosity distance is now easier than fishing in the bucket. We
work in the comoving coordinates and conformal time, i.e, in
Minkowski static space time and then translate back to FRW,
\begin{equation}
\label{lc}
L_c=\frac{hc}{\lambda_c}\frac{dN}{d\eta}=\frac{hc}{\lambda}\frac{dN}{dt}a^2=La^2
\end{equation}
where $L_c$ is comoving Luminosity, $N$ is the total number of
photons radiated by the source at time $t$ and $\lambda_c$ is the
comoving wavelength. For simplicity, we have assumed that all the
photons are emitted with the same wave length. The apparent
brightness is now given by,
\begin{equation}
B=\frac{L_c}{4\pi r^2}= \frac{L}{4\pi d_L^2};~d_L\equiv(1+z)r
\end{equation}
 where $d_L $ is the  luminosity distance in the
expanding universe and $r$ is the comoving distance to the object
from the observer which is constant for a given object but will be
different for different objects. Since, for a photon propagating
radially, $cdt=adr$, the comoving distance can be expressed through
the Hubble parameter
\begin{equation}
\label{dli}
d_L=c(1+z)\int_t^{t_0}{\frac{1}{a(t)}dt}=c(1+z)\int_0^z{\frac{dz'}{H(z')}
};~~\left(\frac{dt}{a}=-\frac{dz}{H(z)}\right)
\end{equation}
which tells that the luminosity distance depends upon the history of
universe. In general, it depends upon the cosmological models.
However, we can extract important information from the above
expression without invoking a particular model. Indeed, for small
$z$, we have,
\begin{equation}
d_L\simeq \frac {c}{H_0}z
\end{equation}
that gives us Hubble law independent of the background cosmological
model which is not surprising as the later is consequence of
homogeneity and isotropy of universe for $v<<c$ or small redshift.
The higher order corrections in $z$ are model dependent. It is
instructive to retain the first order correction to Hubble law(by
invoking the series expansion, $H(z)=H_0+\left({\partial
H}/{\partial z}\right)|_{z=0}z$+.., in (\ref{dli})) ,
\begin{equation}
\label{Q}
 d_L\simeq \frac
{c}{H_0}z\left(1+\frac{1-q_0}{2}z\right);~q_0\equiv-\left(\frac{\ddot{a}}{aH^2}\right)_{t=t_0}
\end{equation}
where $q_0$ is the deceleration parameter which is negative for
accelerating universe as $\ddot a>0$ in this case. It is interesting
to note from Eq.(\ref{Q}) that  the first order correction to Hubble
law distinguishes the accelerating and decelerating models
irrespective of their details such  that $d_L$ is larger in case of
accelerating cosmology. Hence, luminosity distance is an important
indicator of the cosmic history.



Astronomers often quote there observations in terms of distance
modulus defined as,
\begin{equation}
 \mu=m-M=5\log_{10}\left(
 \frac{d_L}{\rm pc}\right)-5 =5\log_{10}\left(\frac{d_L}{\rm Mpc}\right)+25\, ,
 \label{eq:DM}
\end{equation}
where $m$ and $M$ are the apparent and absolute magnitudes of
Supernovae. Distance modulus is the observable which is quoted in
the literature for the Type Ia supernovae observations.

Using Eq.(\ref{eq:DM}) we can express the luminosity distance
through distance modulus,
\begin{equation}
 d_L=10^{0.2\mu+1} ~\rm pc=10^{0.2\mu-5} ~\rm Mpc \, .
 \label{eq:dL1}
\end{equation}

Using then the data from  Type Ia Supernovae (Union2.1
\cite{Suzuki:2011hu} of 580 data points) available for the
observational values of $\mu$, we can calculate the corresponding
values of the luminosity distance from Eq.(\ref{eq:dL1}). We can
also estimate the uncertainty in the luminosity distance data from
the Union2.1  data \cite{Suzuki:2011hu}. Indeed, if we have a
function $f(x,y,...)$ and the errors in $x,y,...$ are given as
$\sigma_x,\sigma_y,...$
 then we can calculate the error in $f$ as,
 \begin{equation}
  \sigma_f^2=\left(\frac{\partial f}{\partial x}\right)^2\sigma_x^2+\left(\frac{\partial f}{\partial y}\right)^2\sigma_y^2+...
  +2\left(\frac{\partial f}{\partial x}\right)\left(\frac{\partial f}{\partial y}\right) \sigma_{xy}+.... \, .
  \label{eq:error}
 \end{equation}
which then gives the uncertainty in $d_L$,
\begin{equation}
 \sigma_{dL}=0.2\ln(10)d_L\sigma_\mu \, ,
\end{equation}
where $\sigma_\mu$ is the error in the measurements of distance
modulus $\mu$.

\begin{figure}[ht]
\begin{center}
\mbox{\epsfig{figure=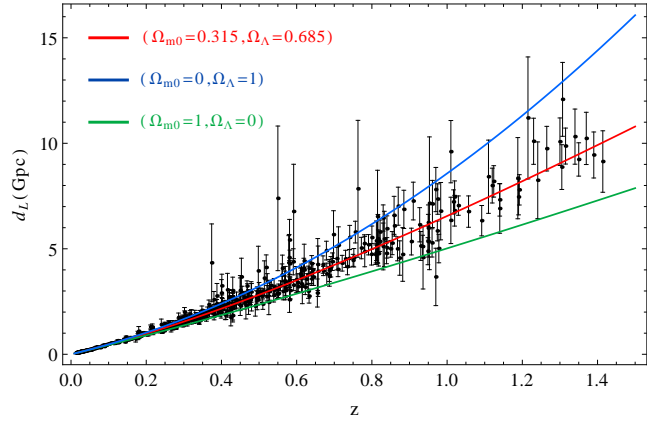,width=8.5cm,angle=0}}
\caption{Luminosity distance is plotted against redshift z for flat
$\Lambda$CDM model. $\Omega_{m0}$ and $\Omega_\Lambda$ in the plot
are the present values of matter and dark energy density parameters.
Black dots are observational values of $d_L$ and black bars are
1$\sigma$ error bars of $d_L$ calculated from Union2.1 data set
\cite{Suzuki:2011hu}.}
\label{fig:dL}
\end{center}
\end{figure}

\begin{figure}[ht]
\begin{center}
\mbox{\epsfig{figure=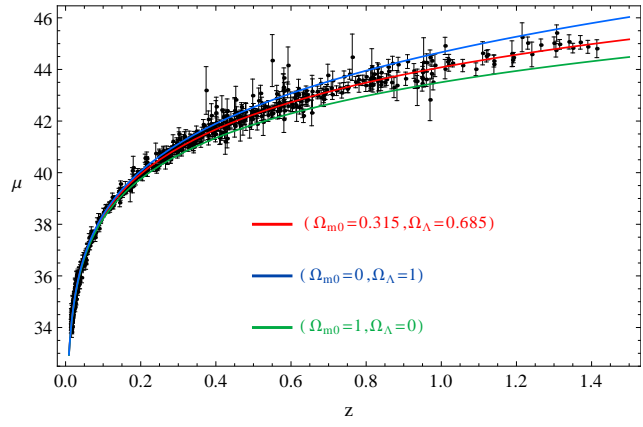,width=8.5cm,angle=0}} \caption{Distance
modulus is plotted against redshift z for flat $\Lambda$CDM model.
$\Omega_{m0}$ and $\Omega_\Lambda$ in the plot are the present
values of matter and dark energy density parameters. Black dots are
observational values of $\mu$ and black bars are 1$\sigma$ error
bars for $\mu$ taken from Union2.1 data set \cite{Suzuki:2011hu}.}
\label{fig:mu}
\end{center}
\end{figure}

In Fig. \ref{fig:dL} we have shown the variation of luminosity
distance with redshift for three different cases of $\Lambda$CDM
model and compared  them with the observational data calculated from
Union2.1 data set \cite{Suzuki:2011hu}. It should be noticed that
error bars are  large at large redshifts. It is clear,however, from
Fig. \ref{fig:dL} that the case $\Omega_{m0}\sim 0.3$ and
$\Omega_\Lambda\sim 0.70$ is best fitted though the other cases
cannot be clearly excluded because of the large error bars of $d_L$
at large redshifts. In Fig. \ref{fig:mu}, we plot distance modulus
$\mu$ versus the redshift and compare with the Union2.1 data set
\cite{Suzuki:2011hu}. As in case of distance modulus, error bars are
 not very large, it is  better seen from Fig. \ref{fig:mu} that the
case $\Omega_{m0}\sim .3$ and $\Omega_\Lambda\sim 0.70$ is best
fitted and others are less  favored.

\subsection{Problem with cosmological constant}

It is clear from the aforesaid that the standard model of Universe
to be consistent with observations should contain a small parameter,
the cosmological constant. It should be noticed that constant energy
plays here a decisive role which is specific to general theory of
relativity though we deliberately kept it out of discussion. For
instance, in mechanics, adding a constant to potential does not
reflect in the underlying physics.

All seems to be well with cosmological constant at classical level.
Problem arises if we bring in here the quantum mechanics
perceptions. Let us try to explain it in simple words. There are
different types of fields which are present in the universe, for
instance the electromagnetic field that we are acquainted with. A
field is a complicated object which is a collection of infinite many
harmonic oscillators each with energy levels given by,
$E_n=(n+1/2)\hbar\omega;~n=0,1,2,...$ such that there is always a
zero point energy, namely, $\hbar\omega/2$ and it is important in
the present context context. When we sum up the zero point energy of
all the oscillators, we obtain the so called vacuum energy, $\rho_{\rm vac}$
which is formally  infinite. Since we know nothing beyond  the
Planck scale, we cut it off there adopting $\rho_{\rm vac}=M^4_p$ as an
expression of our ignorance. This quantity should now be added to
$\rho_\Lambda$ present in equations of motion. The observed value
then is given by,
\begin{equation}
\rho_\Lambda^{obs}=\rho_\Lambda+\rho_{\rm vac}
\end{equation}
We computed $\rho_{\rm vac}$ and  we know $\rho_\Lambda^{\rm obs}$ from
observations but do not in general know $\rho_\Lambda$ which is a
parameter present in the framework at classical level. But since
$\rho_{\rm vac}/\rho_\Lambda^{\rm obs}\sim 10^{120}$, one thing is sure and
certain that the cancelation between $\rho_{\rm vac}$ and $\rho_\Lambda$
should be accurate to one part in $10^{-120}$. We do not know any
mechanism which could give rise to such a fantastic cancelation and
explain as to why this quantity is so small  or why the vacuum
energy gravitates so insignificantly. At low energies, there is no
known symmetry which could give rise to the required cancelation.
This is the famous cosmological constant problem which needs to be
addressed.

\subsection{ The standard candles and their underlying physics in brief  }

For the construction of the Hubble diagram, we need the luminous
objects of known absolute luminosity or standard candles such that
their distance can be found from their apparent luminosity.
Supernovae Ia are best suited as cosmological distance indicators.
Their absolute luminosity can be determined from the physical
conditions responsible for their formation. The type Ia supernovae
form in binary star systems in case one of the companion stars is a
white dwarf with mass below the Chandrasekhar limit $1.4 M_\odot$
and the other is a red giant.

Before we go ahead let us say few words about the life cycle of
stars which crucially depends upon their mass. Stars shine due to
fusion of hydrogen into helium in their core such that the support
against gravity is provided by the thermal pressure. However, as the
stars run out of the hydrogen fuel, their core begins collapsing
under gravitational pressure thereby getting hotter whereas hydrogen
fusion still continue in outer shells. The core which gets hotter
and hotter due to contraction then  pushes the outer layers of the
star outward. As a result the  ejecting layers of the star expand
and cool transforming the star finally into a {\it red giant}. What
happens further crucially depends upon the mass of the core. For
stars with masses in the range of the sun, the ejection process
continues leaving behind a hot core dubbed white dwarf which
eventually cool.
The white dwarf is gravitationally a very compact object typically
with a mass of the order of $M_\odot$ and radius around that of the
earth such that the matter density in the star is given by,
$\rho\sim 10^{12}gm/cc$. It is
 composed
of nuclear waste Carbon and Oxygen supported by the electron
degeneracy pressure alone as its temperature is nearly zero.

 In case the mass  of the collapsing core
of star is between $M_{cs}$ and and $3 M_\odot$, the system ends up
with neutron star. The core with mass more than three solar masses
collapses into black hole.

Let us now discuss the mechanism of supernovae Ia explosion. In the
binary system, the white dwarf begins accreting matter from the
companion star, the reg giant, and gets smaller further. But there
is a limit to this process, as the white dwarf reaches a mass equal
to the Chandrasekhar limit $M_{cs}$, the degeneracy pressure fails
to support the gravitational pressure. Its temperature during
compression increases to the level sufficient for igniting the
carbon fusion which leads to a violent supernovae Ia explosion.

 In what follows, we briefly outline the underlying physics of the explosion. The white
dwarf consists of plasma of carbon-oxygen nuclei and electrons with
temperature around zero kelvin. As the density in white dwarf is
high, some of the electrons are forced to occupy higher momentum
states due to Pauli exclusion principle.  In this case, the electron
pressure dubbed {\it degeneracy pressure}  is of quantum nature
which supports the star against its gravitational pressure. The
electron degeneracy pressure is given by,
\begin{equation}
P_{dg}=\frac{1}{3}\int_0^\infty{v p n(p)dp}
\end{equation}
where (v, p) designate the velocity and momentum of the electron and
$n(p)$ is the number density of electrons per unit momentum
interval. The degeneracy pressure $P_{dg}$ can be computed using
quantum statistical mechanics. In a simpler way, one can  use the
Heisenberg uncertainty relation to estimate it. Indeed, the minimum
volume per electron in the phase space is given by, $\Delta x \Delta
y \Delta z \Delta p_x \Delta p_y \Delta p_z \sim h^3$. The three
dimensional space breaks into small cells of volume $\Delta V=\Delta
x \Delta y \Delta z$. In accordance with Pauli exclusion principle,
this elementary volume can accommodate at most two electrons. The
number density of electrons inside the star is given by, $2/\Delta
V$ and since at low temperature, electrons occupy all the quantum
energy states up to a maximum value $E_F$ dubbed {\it Fermi energy}
with the corresponding momentum $p_F$, we can estimate the total
number of electrons in the white dwarf,
\begin{equation}
\label{npf}
 N_e=\frac{8\pi V}{{h}^3}\int_0^{p_{F}}{p^2 dp}\to
p_{F}=\left(\frac{3h^3N_e}{8\pi V}\right)^{1/3}
\end{equation}
where $V$ is the volume of star.
 The number of electrons in the star
can be expressed through its density $\rho$,
\begin{equation}
N_e=V\frac{\rho Z}{M_p A} \sim
\frac{\rho}{M_p},~\left(\frac{Z}{A}\simeq \frac{1}{2}\right)
\end{equation}
where Z(A) is atomic number(mass number) and $M_p$ is the mass of
proton. In case the mass of white dwarf is much smaller than
$M_{cs}$, electrons in the white dwarf can be treated non
relativistically ($v<<c$) such that $v\simeq p/m_e$,
\begin{equation}
\label{pe}
P_{dg}=\frac{8\pi}{3m_eh^3}\int_0^{p_{F}}{p^4dp}=\frac{8\pi}{15m_eh^3}p_{F}^5=k_1\rho^{5/3};
~k_1=\left(\frac{3h^3}{8\pi}\right)^{2/3} \frac{1}{5m_e M_p^{5/3}}
\end{equation}
where $\rho$ is matter density in the white dwarf. Let us note that
the degeneracy pressure is inversely proportional to the mass of
particles thereby the pressure due to nucleons is negligible as
compared to the pressure caused by electrons. Next, we need to
compare the degeneracy pressure with gravitational pressure of the
white dwarf. The self gravitational energy or the binding energy of
the white dwarf is given by,
\begin{equation}
E_g=-\frac{3}{5}\frac{GM^2}{R}
\end{equation}
where $M$ is  mass and $R$ is radius of the star.  The corresponding
gravitational pressure is given by the thermodynamic expression,
\begin{equation}
P_g=-\frac{\partial}{\partial
V}E_g=-\left(\frac{4\pi}{375}\right)^{1/3}GM^{2/3}\rho^{4/3}
\end{equation}
The degeneracy pressure encounters the gravitational pressure of the
white dwarf and equating the two we obtain,
\begin{equation}
k_1\rho^{5/3}=\left(\frac{4\pi}{375}\right)^{1/3}GM^{2/3}\rho^{4/3}\to
R \sim \frac{1}{M^{1/3}}
\end{equation}
Thus the volume of white dwarf is inversely proportional to its
mass. Consequently, as white dwarf gets heavier during accretion
process, its gets smaller and thereby denser. The electrons then are
forced to occupy still higher momentum states such that $p_F$ is
large. The non relativistic approximation then ceases to apply. In
this case we should use the relativistic expression for velocity,
\begin{equation}
v=\frac{\frac{p}{m_e}}{\sqrt{ 1+(\frac{p}{m_ec})^2 }  }
\end{equation}
 while computing the degeneracy
pressure. In this case, it is possible to compute the integral in
(\ref{pe}) in the closed form. It would be instructive to quote the
expression of $P_{dg}$ in the limit $p/mc>>1$,
\begin{equation}
\label{per}
 P_e{dg}\simeq
k_2\rho^{4/3}-k_3\rho^{2/3};~k_2={c}\left(\frac{3h^3}{8\pi M_p^4}
\right)^{1/3};~k_3={c^3}\left( \frac{\pi
m_e^2}{24h^3M_p^2}\right)^{1/3}
\end{equation}
The first leading term corresponds to the ultra-relativistic case
($v\simeq c$) and the second term provides the first order
correction. Equating (\ref{per}) to the gravitational pressure, we
arrive at an important relation,
\begin{equation}
\label{r}
 R\sim
M^{1/3}\sqrt{1-\left(\frac{M}{M_{ch}}\right)^{2/3}};~M_{ch}=\left(\frac{5k_2}{G}\right)^{3/2}\left(\frac{3}{4\pi}\right)^{1/2}\simeq
1.7 M_\odot
\end{equation}
\begin{figure}[h]
\label{sp2}
\includegraphics[scale=0.6]{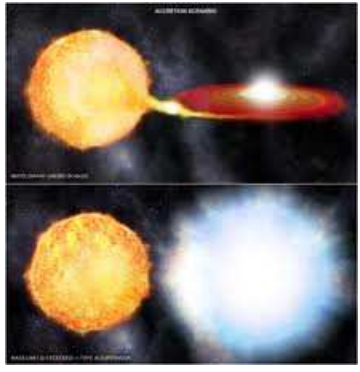}
\caption{ Figure shows a binary system with a white dwarf and
companion red giant. In the upper panel, the white dwarf accretes
mass from the outer layer of the red giant and shrinks. As the mass of
white dwarf reaches the Chandrasekhar limit, the Carbon fusion
ignites leading to supernova Ia explosion shown in the lower panel
of the figure. Image: NASA/CXC/M Weiss }
\end{figure}

\begin{figure}[h]
\label{sp3}
\includegraphics[scale=0.8]{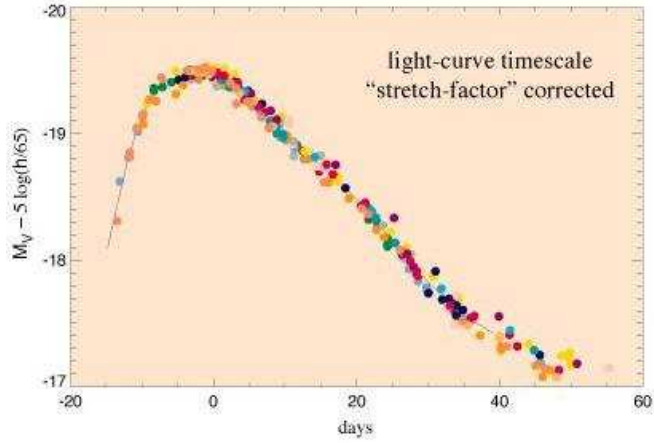}
\caption{ Supernova Ia light curve shows the variation of relatives
brightness versus the days after explosion. The figure is taken from the site
\url{www-supernova.lbl.gov}.}
\end{figure}
It is interesting to note that if we drop the correction to the
leading term in Eq.(\ref{per}), i.e., if we stick to the
ultra-relativistic limit and then compare the degeneracy pressure to
the gravitational pressure, the radius dependence drops out giving
rise to,
\begin{equation}
M=\left(\frac{5k_2}{G}\right)^{3/2}\left(\frac{3}{4\pi}\right)^{1/2}\equiv
M_{ch}
\end{equation}
which tells that $M_{ch}$ is the limiting value of the mass and
Eq.(\ref{per}) tells us how radius of the star depends upon its mass
as $M\to M_{ch}$. In order to get better feeling for the limiting
mass of the white dwarf let us show that the said limit is related
to the stability of the star. In the relativistic approximation, the
kinetic energy of the degenerate electron gas in the star is given
by,
\begin{equation}
E_K\simeq c p_F\simeq \frac{hc N_e^{1/3}}{R}\simeq \frac{hc
N_p^{1/3}}{R}
\end{equation}
where $N_p$ is the number of baryons in the star and $p_F$ is given
by Eq.(\ref{npf}). Thus, the total energy of the white dwarf can be
estimated as,
\begin{equation}
E_{tot}\simeq \frac{hc N_p^{1/3}}{R}-\frac{G N_p M_p^2}{R}
\end{equation}
Stability then demands that the maximum number of baryons the white
dwarf can accommodate should correspond to $E_{tot}=0$,
\begin{equation}
N_p^{max}\simeq \left(\frac{hc}{GM_p^2}\right)^{3/2}\to M_{ch}\simeq
 M_p\left(\frac{hc}{GM_p^2}\right)^{3/2}
\end{equation}
which is the same order of magnitude we obtained by comparing the
degeneracy pressure with the gravitational pressure.\\
 Our estimates depend upon the
assumption we made for simplicity that matter density inside the
star is constant. The realistic calculations which incorporate
density variation inside the white dwarf,  give, $M_{ch}\simeq 1.44
M_\odot$. Our estimate formally tells us that as $M\to M_{ch}$, the
radius goes to zero.


Actually, as the mass of the white dwarf increases, temperature
inside the star rises due to compression where as the volume is
still controlled by the degeneracy pressure. The temperature  in the
core of the star increases to the level that Carbon fusion ignites
which further heats it up. The white dwarf can not expand and cool
like an ordinary star as its volume is controlled by the degeneracy
pressure and the thermal pressure is negligible, contraction in this
case continues till the thermal energy remains below  Fermi energy.
As a result, temperature rises to the extent that  thermal
conduction can no longer cope up with nuclear burning giving rise to
run away nuclear reactions. What happens thereafter is a topic of an
active debate at present. However, there is a broad consensus that
as runaway nuclear reaction ensues, the  matter in the star quickly
fuses into iron-peak elements releasing enormous amount of energy
giving rise to
supernova explosion about 5 billion times brighter than the Sun. The
luminous event can be identified by its light curve which shows the
rapid increase of luminosity to a maximum value and then disappears
in one to two months time. The last supernova in our galaxy was seen
in 1604; around 300 supernovae are seen from other galaxies every
year. Since these events are rare, their systematic search using
their light curves needs to be carried out.

Based upon the aforesaid discussion, we draw the following important
conclusion. As the underlying physics mechanism for supernovae Ia
explosions is common or {\it all the supernovae Ia form under the
same physical conditions, their absolute luminosity is same}. Hence,
the supernovae which look dimmer are farther deeper in the universe.
The observational investigations of their luminosity distances
provides important information of the cosmic history.


\section{Discussion and outlook}
We described the kinematics and dynamics of homogeneous and
isotropic universe in the framework of Newtonian cosmology. Since
the material is prepared for a wider audience, we often resorted to
heuristic arguments to convey the underlying physics ideas. We
mentioned shortcomings of the standard model of Universe and
focussed on the late time inconsistency dubbed age crisis in the hot
big bang and argued that accelerated expansion plays an important
role in the history of universe. In the standard framework, in
particular, the resolution of age problem requires a repulsive
effect or late time acceleration which dominates over deceleration
due to normal matter. The repulsive effect can be caused by a
positive cosmological constant in the standard lore or by the large
scale modification of gravity. We tried to impress upon that similar
to the early universe problems related to flatness, horizon and
primordial fluctuations whose resolution requires an early phase of
accelerated expansion, the solution of age problem also asks for the
late time cosmic acceleration.

Cosmological constant, in a sense, is the simplest device that can
turn gravity into a repulsive force at late times. It is remarkable
that the effect was directly seen in supernovae Ia observations in
1998. The repulsive effect can also be caused by the slowly rolling
scalar fields or by a large scale modification of gravity. We
briefly explained how a tiny mass of graviton could mimic
cosmological constant like behavior.

 We also included a simple
discussion on the underlying physics of supernovae Ia
explosions and  their role of distance indicators for the
understanding of cosmic history. The luminosity distance is larger
in case of accelerating models such that the related supernovae Ia
data clearly speaks in favor of late time cosmic acceleration which
is indirectly supported by other observations. The early phase of
cosmic acceleration {\it a la} inflation fixes the total cosmic
energy budget out of which nearly 30 $\%$ is contributed by cold
dark matter as revealed by the study of large scale structures and
the missing 70 $\%$ is fixed by the observed late time acceleration
of Universe.

At present, observations are not in position to distinguish between
various options that might give rise to late time acceleration. The
future surveys might unveil the true cause of the phenomenon. It is
quite likely that there is nothing but $\Lambda$ which is
responsible for late time acceleration. Neither the scalar fields
nor any known scheme of large scale modification perform better than
the cosmological constant both on theoretical and observational
grounds. However, non linear massive gravity, despite several tough
theoretical challenges it faces, still deserves attention. It links
cosmological constant to the mass of a fundamental particle, the
graviton and it has an in built mechanism of de gravitation which is
a novel perspective for addressing the cosmological constant
problem. {\it Cosmological constant problem is the problem} which
needs to be finally addressed.

Whatever may be the underlying reason, we need a dominant
(effective) dark energy component at late times to support the old
universe we live in. If the Universe is to grow that old it is
today, it has to be predominantly dark which sounds like an
Anthropic argument!.
\section{Acknowledgements} I am grateful to  J.V. Narlikar for his kind
comments. I also thank Wali Hossain  and Shahalam for useful
discussions and for drawing some of the figures included in the
text.

\end{document}